\documentclass[aps,prb,twocolumn,floatfix,amssymb,amsfont,amsmath,float]{revtex4-1}

\usepackage{mathptmx}
\usepackage{subfigure}
\usepackage{dcolumn}
\usepackage{amsmath,amssymb}
\usepackage{bm}
\usepackage{color}
\usepackage{latexsym}
\usepackage[english]{babel}
\usepackage{times}
\usepackage{wasysym}

\usepackage{psfrag,graphicx}
\usepackage{epsf}
\usepackage{amsfonts}
\usepackage{natbib}
\usepackage{appendix}
\usepackage[colorlinks=true,citecolor=blue,linkcolor=blue]{hyperref}

\begin{document}

\title{Devil's staircases without particle-hole symmetry}

\author{Zhihao Lan, Igor Lesanovsky, and Weibin Li}
\affiliation{School of Physics and Astronomy, University of Nottingham, Nottingham NG7 2RD, United Kingdom}
\affiliation{Centre for the Mathematics and Theoretical Physics of Quantum Non-equilibrium Systems, University of Nottingham, Nottingham NG7 2RD, United Kingdom}
\begin{abstract}
\textbf{}
We present and analyze spin models with long-range interactions whose ground state features a so-called devil's staircase and where plateaus of the staircase are accessed by varying two-body interactions. This is in contrast to the canonical devil's staircase, for example occurring in the one-dimensional Ising model with long-range interactions, where typically a single-body chemical potential is varied to scan through the plateaus. These systems, moreover, typically feature a particle-hole symmetry which trivially connects the hole part of the staircase (filling fraction $f\geq1/2$) to its particle part ($f\leq1/2$). Such symmetry is absent in our models and hence the particle sector and the hole sector can be separately controlled, resulting in exotic hybrid staircases.
\end{abstract}

\maketitle

\section{Introduction}
A devil's staircase is a fractal structure that characterizes the ground state of a plethora of systems in physics \cite{bak_review, bak_physics_today}. Examples include the Frenkel-Kontorowa model \cite{Frenkel_Kontorova_book, staircase_aubry}, the Falicov-Kimball model \cite {staircase_2D_Falicov_Kimball, staircase_1D_Falicov_Kimball, staircase_Falicov_Kimball1D2D}, Ising models \cite{staircase82, bak_annni}, quantum dimer models \cite{staircase_QDM1, staircase_QDM2, staircase_QDM3} as well as certain discrete maps \cite{staircase_Circle_Map, staircase_double, staircase_multiple}. In particular, one-dimensional (1D) Ising models are paradigmatic systems that may exhibit devil's staircases both in the case of long-range and short-range interactions. In Ising models with long-range interactions it was shown rigorously that the permitted filling fractions (ratio of the number of particles to that of lattice sites) of the ground state configurations form a complete devil's staircase when the chemical potential is varied \cite{staircase82}. This means when scanning the chemical potential the filling fractions can assume all rational numbers. For the short-range interacting anisotropic next-nearest-neighbour Ising (ANNNI) model, such staircase appears only at finite temperatures \cite{bak_annni} since there are solely two stable ground states (known as ferromagnetic phase and antiphase) at zero temperature. The staircase structure of ANNNI model has been observed in NaV$_2$O$_5$ under high pressure \cite{staircase_NaV2O5}.

In recent years, controllable quantum systems have emerged as platforms for exploring phenomena in condensed-matter and high-energy physics \cite{quantum_simulation}. This includes trapped ions \cite{ion_staircase1, ion_staircase2}, cold polar molecules \cite{molecular_staircase1, molecular_staircase2, molecular_staircase3, molecular_staircase4, molecular_staircase5} and strongly interacting Rydberg atoms \cite{Rydberg_staircase1, Rydberg_staircase2, Rydberg_staircase3, Rydberg_staircase4, Rydberg_staircase_gs1, Rydberg_staircase_gs1, RydCry_science}, ultracold atoms in bichromatic lattices \cite{optical_latice_staircase}, synthetic dimensions and gauge fields \cite{SynDim}, photons with engineered long-range interactions \cite{photonic_staircase} and optomechanical cavity systems \cite{optomechanical_staircase}. These platforms allow not only to control single-body quantities (e.g. the trapping potential or the chemical potential), but also to tailor the shape of the underlying two-body interaction.

In a recent study~\cite{staircase15PRL}, we have identified a new mechanism underlying the formation of a devil's staircase within a spin model implemented by Rydberg atoms held in a 1D optical lattice. By using a so-called double-dressing scheme~\cite{staircase15PRL}, we have shown how to create competing interactions with short-range attraction and longer-range repulsion between two atoms. In particular, we focused on a situation where the nearest-neighbour interaction is attractive and tunable while the interactions from next-nearest-neighbours onwards follow a repulsive van der Waals (vdW) potential. Such non-convex potential leads to the formation of a devil's staircase in the ground state and its plateaus are accessed by varying the strength of the nearest-neighbour attraction.

This situation is in contrast to that encountered, e.g. in above-mentioned staircase of the Ising model, which is governed by a single-body chemical potential term. The staircase in the Ising model features a particle-hole symmetry, i.e., the hole part of the staircase at filling fraction $f\geq1/2$ can be trivially extracted from the particle part of the staircase at $f\leq1/2$. In our previous work~\cite{staircase15PRL}, we have shown that a broken particle-hole symmetry emerges for a staircase whose plateaus are accessed by two-body attractive interactions. In this situation the staircase is a union of two sub-staircases that are consisting of either dimer particles or dimer holes. This finding opens up the possibility to study a plethora of hybrid staircases. For example, it is possible to encounter a situation with a dimer particle sub-staircase in the particle sector and a trimer hole sub-staircase in the hole sector. Finally, we would like to note that the impact of different kinds of interactions on the devil's staircase physics has been studied in the literature from different perspectives \cite{R1, R2, R3, R4, R5, R6, R7, R8}. For example, some aspects of the staircases discussed in the present work can be linked to studies of atoms adsorbed on a surface \cite{R3, R5}. However, a systematic exploration of devil's staircases from the perspective of particle-hole symmetry breaking has not been conducted previously. 

In this work, we extend our previous study \cite{staircase15PRL} to spin models that feature attractive interactions not only among nearest neighbours but over a longer range. The paper is structured as follows: In section~\ref{sec2}, we present the model Hamiltonian and discuss the role played by the particle-hole symmetry. In section~\ref{sec3}, we discuss analytical and numerical tools for analyzing the ground state properties of the model Hamiltonian. In section~\ref{sec4}, we benchmark our tools by applying them to a conventional staircase controlled by a chemical potential. In section~\ref{sec5} we investigate in detail the situation where two-body interactions drive staircases without particle-hole symmetry, which constitutes the central part of this work. In section~\ref{sec6}, we discuss the possibility of staircases controlled by $n$-body interactions ($n>2$). We conclude and provide an outlook in section~\ref{sec7}.

 %%%%%%%%%%%%%%%%%%%%%%%%%%%%%%%%%%%%%%%%
\begin{figure}[htbp]
	\centering
	\includegraphics[width=0.8\columnwidth]{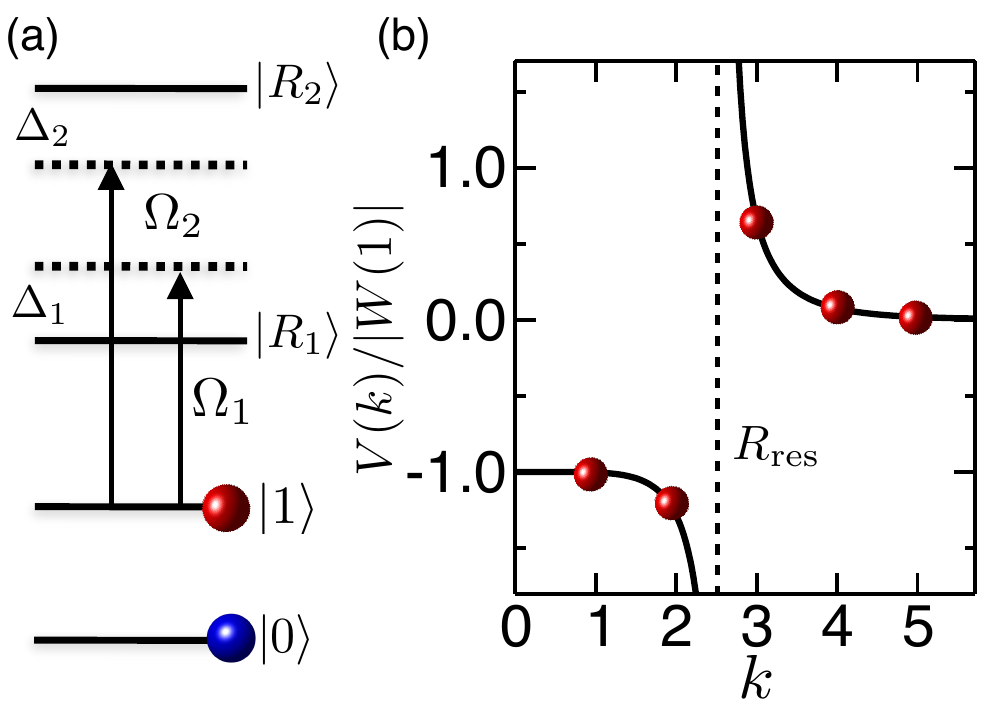}
	\caption{(a) Level scheme. An electronically low-lying state $|1\rangle$ is laser-coupled to Rydberg states $|R_1\rangle$ and $|R_2\rangle$ with Rabi frequency $\Omega_1$ and $\Omega_2$, and detuning $\Delta_1$ and $\Delta_2$ respectively.  (b) Effective interaction potential between particles in dressed state $|1\rangle$. This interaction is attractive at short distances and repulsive at long distances. Here, we show a situation where the nearest neighbour and next-nearest neighbour are attractive, i.e. $W(1)<0$ and $W(2)<0$.  }
	\label{fig:level}
\end{figure}

\section{Model Hamiltonian}
\label{sec2}
Staircases explored in this work rely on a non-convex long-range interaction which is attractive at short distances and repulsive at large distances. In our previous work~\cite{staircase15PRL},  we proposed that a special form of this interaction, i.e., with van der Waals repulsive tail, can be engineered with the help of Rydberg atoms. Specifically the physical setting is a 1D lattice with spacing $a$ where each site can either be occupied by an atom in state $|1\rangle$ or $|0\rangle$. For convenience, we denote that a site is empty (occupied by a particle) when the atom of the site is in state $|0\rangle$ ($|1\rangle$). We then employ a double-dressing scheme~\cite{staircase15PRL}, in which two blue- and red-detuned lasers are applied simultaneously to weakly couple the $|1\rangle$ state with two Rydberg $S$-states $|R_1\rangle$ and $|R_2\rangle$, as depicted in Fig.~\ref{fig:level}(a). The Rabi frequency and detuning of the blue (red) detuned laser are $\Omega_1$ ($\Omega_2$) and  $\Delta_1$ ($\Delta_2$), respectively. The vdW interaction of the Rydberg state $|R_j\rangle$ is $C_j/r^6$ with $C_j$ the corresponding dispersion constant ($j=1,2$). We will neglect the inter-state Rydberg interaction when the two states are far separated energetically \cite{Olmos11}. The lasers induce long range interactions between atoms in the Rydberg dressed $|1\rangle$ state \cite{Henkel10PRL, Honer10PRL, Cinti10PRL, Li12PRA}. The blue-detuned laser induces an interaction potential $U_1(r)=\tilde{C}_1/(r^6-R_{\text{res}}^6)$, where $\tilde{C}_1=R_{\text{res}}^6\Omega_1^4/8\Delta_1^3$ and $R_{\text{res}}=(C_1/2|\Delta_1|)^{1/6}$ determines the distance of the two-atom resonant excitation when $2\Delta_1+C_1/R^6_{\text{res}}=0$ \cite{Ates12PRL, Li13PRL}. The resulting interaction is attractive for $r<R_{\text{res}}$ and repulsive when $r>R_{\text{res}}$. The red-detuned laser generates a long range soft-core interaction, $U_2(r)=\tilde{C}_2/(r^6+r_2^6)$ where $\tilde{C}_2=r_2^6\Omega_2^4/8\Delta_2^3$ and the core radius $r_2=(C_2/2|\Delta_2|)^{1/6}$. The overall dressed interaction is given by the combined potential of $V(r)=U_1(r) + U_2(r)$, which is illustrated in Fig.\ref{fig:level}(b). By tuning the laser parameters, strength and attractive range of the non-convex long-range two-body interactions can be varied (details of the implementation is given in Ref. [35]).

In this work, we will go beyond this special realization with Rydberg atoms and consider more general  non-convex interactions, where the repulsive tail is not limited by the vdW type, i. e. $V(r)\sim 1/r^{\alpha}$ with $1<\alpha$, focusing more on the physics rather than the experimental implementations. When $\alpha$ is taken as a parameter that can be freely tuned, many new features are found in the respective staircase which is not revealed using the vdW interaction. Taking these considerations into account, we study a classical 1D spin chain governed by the following Hamiltonian
\begin{eqnarray}
H=\sum_{i=-\infty}^{\infty}  \sum_{r=R+1}^{\infty}V(r) n_i n_{i+r}+\sum_{i=-\infty}^{\infty} \sum_{r=1}^{R} W(r) n_in_{i+r},
\label{2bHp}
\end{eqnarray}
\noindent
where $W(r)\le 0$ ($r=1, \cdots, R$) parametrizes the strength of the attractive potential part, with $R$ to be the range of the attractive interaction. The potential $V(r)=(R+1)^{\alpha}/r^{\alpha}$ ($r=R+1, \cdots$) corresponds to the repulsive tail. Note that here and in the following, the energy is expressed in units of $V({R+1})$ and length in units of the lattice spacing $a$.

A particle-hole symmetry is absent in Hamiltonian~(\ref{2bHp}) which is explicitly seen by applying the particle-hole transformation, $n_i=1-m_i$, where $m_i$ denotes the occupation of a hole at $i$-th site. This yields the Hamiltonian for the holes,
\begin{eqnarray}
H&&=\sum_{i=-\infty}^{\infty}  \sum_{r=R+1}^{\infty}V(r) m_i m_{i+r}+\sum_{i=-\infty}^{\infty} \sum_{r=1}^{R} W(r) m_im_{i+r}\nonumber\\
&&-\mu' \sum_{i=-\infty}^{\infty} m_i+C,
\label{2bHh}
\end{eqnarray}
where $\mu'= 2\sum_{r=R+1}^{\infty} V(r)+2\sum_{r=1}^{R} W(r)$ and $C=\sum_{i=-\infty}^{\infty}  \sum_{r=R+1}^{\infty}V(r)+\sum_{i=-\infty}^{\infty} \sum_{r=1}^{R} W(r)$. The extra $\mu'$ term, which is controlled by interactions $V(r)$ and $W(r)$, is typically nonzero. In this case, the Hamiltonian of the hole is structurally different from that of the particle. 

\section{Methods}
\label{sec3}
To investigate the ground state of Hamiltonian (\ref{2bHp}), we will use both analytical and numerical tools. The analytical method is based on that by Bak and Bruinsma~\cite{staircase82}. It was originally used to deal with repulsive and convex interactions and we will adapt it to our system. The analytic treatment is accompanied by ``brute-force" numerical calculations to find the ground state of (\ref{2bHp}).

\subsection{Analytical method}

\subsubsection{Stability regions of monomers}
When studying the Ising model with convex interactions, Bak and Bruinsma \cite{staircase82} showed that for any rational filing fraction $f=\frac{q}{p}$ ($p$ and $q$ are nonnegative integers) of the particles to the lattice sites, there will be a finite range, i.e., a stability region of chemical potential (with lower and upper bound $\mu^-$ and $\mu^+$, respectively) such that the most homogeneous configuration with this filling fraction is the ground state configuration.

The stability regions are determined by the following equations (the derivation is for convenience given in Appendix~\ref{secA}),
\begin{gather}
\mu^{-}=\sum_{n=1, \alpha_n\neq 0}^{\infty} [(r_n+1)V(r_n)-r_nV(r_n+1)] \nonumber \\
+\sum_{n=1, \alpha_n=0}^{\infty}[(r_n+1)V(r_n)-r_n V(r_n+1)]
\label{mu1}
\end{gather}
and
\begin{gather}
\mu^{+}=\sum_{n=1, \alpha_n\neq 0}^{\infty} [(r_n+1)V(r_n)-r_nV(r_n+1)] \nonumber \\
+\sum_{n=1, \alpha_n=0}^{\infty}[(-r_n+1)V(r_n)+r_nV(r_n-1)]
\label{mu2}
\end{gather}
where $r_n$ and $ \alpha_n$ are related to $p$ and $q$ through the relation $np=r_nq+\alpha_n$ with $0\leq \alpha_n<q$. From these equations, we obtain the ``width" of the stability region,
\begin{eqnarray}
\Delta_\mu&=&\mu^{+}-\mu^{-} \nonumber \\
&=&\sum_{n=1, \alpha_n=0}^{\infty}[r_nV(r_n-1)-2r_n V(r_n)+r_nV(r_n+1)].
  \label{muT}
\end{eqnarray}

\subsubsection{Effective interaction between two n-mer particles and holes}
In our case particles tend to form clusters due to the short-range attraction. In general, if the first $R$ nearest-neighbour interactions are attractive, then $R+1$ particles will form a cluster on sites $(i, i+1, \cdots, i+R)$. We will refer to such n-particle (hole) cluster as n-mer particle (hole).
\begin{figure}[htbp]
\centering
\includegraphics[width=0.6\columnwidth]{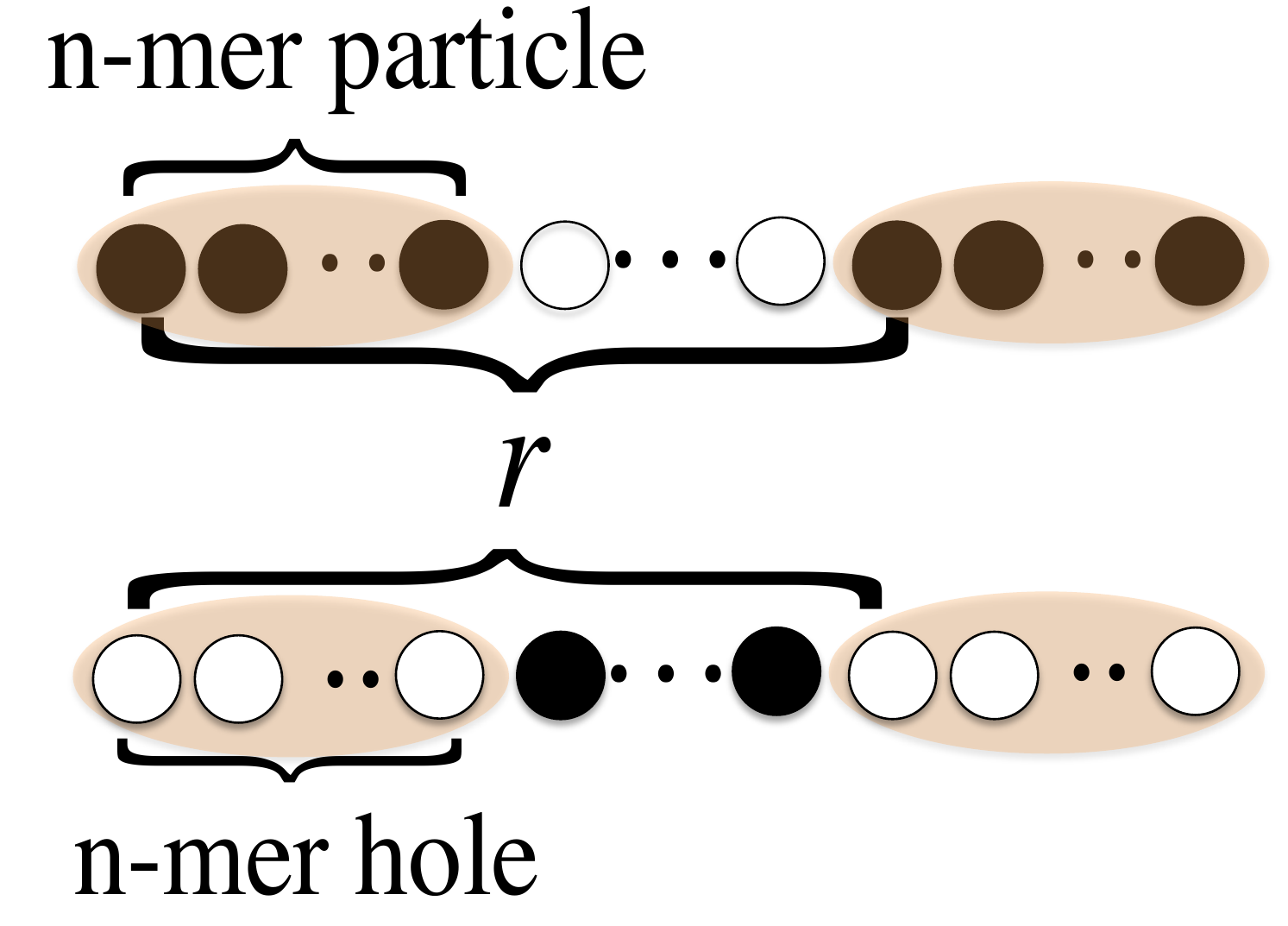}
\caption{Two n-mer particles (holes) separated by distance $r$ will interact according to the 2n binary interactions of their constituent particles (holes), resulting in the interaction matrix given by Eq. (\ref{nmer_interact}).}
\label{fig_nmer}
\end{figure}

The method by Bak and Bruinsma is extended to capture this case by treating an n-mer as an effective ``monomer". To this end one needs to know the effective chemical potential for a n-mer and the interaction between two n-mers. (Note that, for an n-mer with filling fraction $q/p$, the corresponding filling fraction of the actual monomers is $nq/p$.) The interaction between two n-mer particles (holes) separated by $r$ lattice sites, as shown in Fig.~\ref{fig_nmer}, can be conveniently described by a matrix
\begin{eqnarray}
\tilde{V} =
 \begin{pmatrix}
  V(r) &  V(r+1) & \cdots &  V(r+n-1) \\
   V(r-1) &  V(r) & \cdots &  V(r+n-2) \\
  \vdots  & \vdots  & \ddots & \vdots  \\
   V(r-n+1) &  V(r-n+2)& \cdots &  V(r)
 \end{pmatrix},
 \label{nmer_interact}
\end{eqnarray}
where matrix element $\tilde{V}_{ij}$ describes the interaction between the $i$-th  particle of the first $n$-mer and the $j$-th particle of the second $n$-mer. The effective interaction between two n-mers is then given by $\tilde{V}_{\rm{eff}}(r)=\sum_{ij} \tilde{V}_{ij}$.  Under the condition that  there is no overlap of two n-mers in the most homogeneous configuration, we can use the method by Bak and Bruinsma to describe the n-mers, when we replace the interaction of Eqs. (\ref{mu1}), (\ref{mu2}), and  (\ref{muT}) by $\tilde{V}_{\rm{eff}}(r)$. Furthermore, we need to replace the monomer chemical potential by that of the n-mer particle or hole, which will be discussed in the following.

\subsubsection{Effective chemical potential of n-mer particle and hole}
In case of the n-mer particle, we find that $n$ particles will cluster together to lower their energy when the range of the attractive interactions is $R=n-1$. According to Hamiltonian (\ref{2bHp}), the interaction energy within the n-mer particle then serves as effective chemical potential given by
\begin{gather}
\mu_{np}=W(n-1)+2W(n-2)+\cdots+(n-1)W(1).
\end{gather}
Note, that if  $W(i)=W$ for all $i$, then $\mu_n^p=n(n-1)W/2$.

For a m-mer hole the calculation of the effective chemical potential is more involved as the size of the hole cluster can be larger than the range of the attractive interactions. Using Hamiltonian (\ref{2bHh}), we find the effective chemical potential of a m-mer hole,
\begin{eqnarray}
\label{CPnHole}
\mu_{mh}^{(R)}&&=-E_{mh}^{(R)}=-\sum_{j=1}^{R}(m-j)W(j)-\sum_{k=R+1}^{m-1} (m-k)V(k)\nonumber \\
&&+2m\sum_{r=R+1}^{\infty}V(r)+2m\sum_{r=1}^RW(r).
\end{eqnarray}
When $m\geq R+1$ the size $m$ depends on both $R$ and the long-range repulsive tail. This is a manifestation of the particle-hole symmetry breaking, i.e., the size of the hole cluster is not necessarily the same as the particle cluster which would give $m=R+1$. In the following, we list explicitly the effective chemical potentials of particle and hole clusters of different sizes for $R=1$ and $R=2$, which are relevant for our discussions below.

\begin{itemize}
\item $R=1$:  the chemical potentials of particle and hole dimers, and hole trimers are given by,
\begin{gather}
\mu_{2p}=W(1) \label{EqR11}\\
\mu_{2h}^{(1)}=4\sum_{r=2}^{\infty}V(r)+3W(1)  \label{EqR12}\\
\mu_{3h}^{(1)}=6\sum_{r=2}^{\infty}V(r)-V(2)+4W(1)  \label{EqR13}
\end{gather}
\item $R=2$:  the chemical potentials of particle and hole trimers, and hole tetramer are given by,
\begin{gather}
\mu_{3p}=W(2)+2W(1) \label{EqR21}\\
\mu_{3h}^{(2)}= 6\sum_{r=3}^{\infty}V(r)+4W(1)+5W(2) \label{EqR22}\\
\mu_{4h}^{(2)}=8\sum_{r=3}^{\infty}V(r)-V(3)+5W(1)+6W(2) \label{EqR23}
\end{gather}
\end{itemize}

In order to obtain the stability regions of n-mer particles and holes [for Hamiltonians (\ref{2bHp}) and (\ref{2bHh})], we can now use Eqs. (\ref{mu1}-\ref{muT}) with the effective interaction $\tilde{V}_{\rm{eff}}(r)$, the effective chemical potentials $\mu_{np}$ and $\mu_{nh}^{(R)}$ as well as the true monomer filling fraction $nq/p$ (associated with a n-mer particle) or hole filling fraction $q/p$.

The above effective theory for n-mer particles and holes only works when the staircase contains no mixtures of n-mers of different kinds. The aim of this work is to understand when the staircase can be described by a union of two pure sub-staircases in the particle and hole sectors, respectively. 

\subsection{Numerical method}

The filling fraction associated with the ground state configuration of Hamiltonian (\ref{2bHp}) as functions of the attractive interaction $W(r)$ ($r=1,2,\cdots, R$) can be calculated by a brute force method. In this numerical method~\cite{staircase15PRL}, we check all possible periodic configurations of an infinite chain with period $p$ up to a certain limit ($p=23$ in this study, due to the limitation of computational resources). The ground state configuration is determined by the one that has the lowest energy density (energy of a single period divided by the length of the periodicity). This captures the coarse structure of the staircase as the phases with large $p$ usually have very small \cite{bak_review} stability regions.

 %%%%%%%%%%%%%%%%%%%%%%%%%%%%%%%%%%%%%%%%
\section{Particle-hole symmetry in traditional devil's staircases}
\label{sec4}
To provide some context, we review here briefly the results by Bak and Bruinsma~\cite{staircase82}, which are based on an Ising model with long-range interactions,
\begin{gather}
H=\sum_{i=-\infty}^{\infty}  \sum_{r=1}^{\infty}V(r) n_i n_{i+r}-\mu \sum_{i=-\infty}^{\infty} n_i,
\label{1bHp}
\end{gather}
\noindent
where $n_i=0,1$ when the site $i$ is empty or occupied by a particle, respectively. Here, $V(r)$ describes a long-range repulsive interaction between two particles separated by $r$ sites and $\mu$ is the chemical potential for the particle. For any rational filling fraction $f$ of the particles, the ground state configuration will assume a distribution in space as uniform as possible if the infinite-range interaction $V(r)$ is strictly convex~\cite{Hubbard, Uimin}. In this case the ground state configuration is independent of the actual details of the interaction potential and features so-called generalized Wigner crystals. The filling fractions $f$ of the ground state configurations form a complete devil's staircase as a function of the chemical potential $\mu$~\cite{staircase82}.

For power-law interactions $V(r)=1/r^{\alpha}$, the Hamiltonian (\ref{1bHp}) is invariant (apart from an irrelevant constant term) under the particle-hole transformation, $n_i=1-m_i$ and the corresponding hole Hamiltonian reads
\begin{gather}
H=\sum_{i=-\infty}^{\infty}  \sum_{r=1}^{\infty}V(r) m_i m_{i+r}-\mu' \sum_{i=-\infty}^{\infty} m_i+C,
\label{1bHh}
\end{gather}
where $\mu'= 2\sum_{r=1}^{\infty} V(r)-\mu$ and $C=\sum_{i=-\infty}^{\infty}  \sum_{r=1}^{\infty} V(r)-\mu\sum_{i=-\infty}^{\infty}$. One can find the transition point to the state without holes (or a state where the lattice is fully occupied by particles) by setting $\mu'=0$, i.e., $\mu=2\sum_{r=1}^{\infty}V(r)$. For power-law interactions, the corresponding critical chemical potential $\mu_c$ is determined by $\mu_c=2\zeta (\alpha)$ with $\zeta(\alpha)=\sum_{n=1}^{\infty}1/n^{\alpha}$ being the Riemann zeta function.

\begin{figure}[htbp]
	\centering
	\includegraphics[width=0.8\columnwidth]{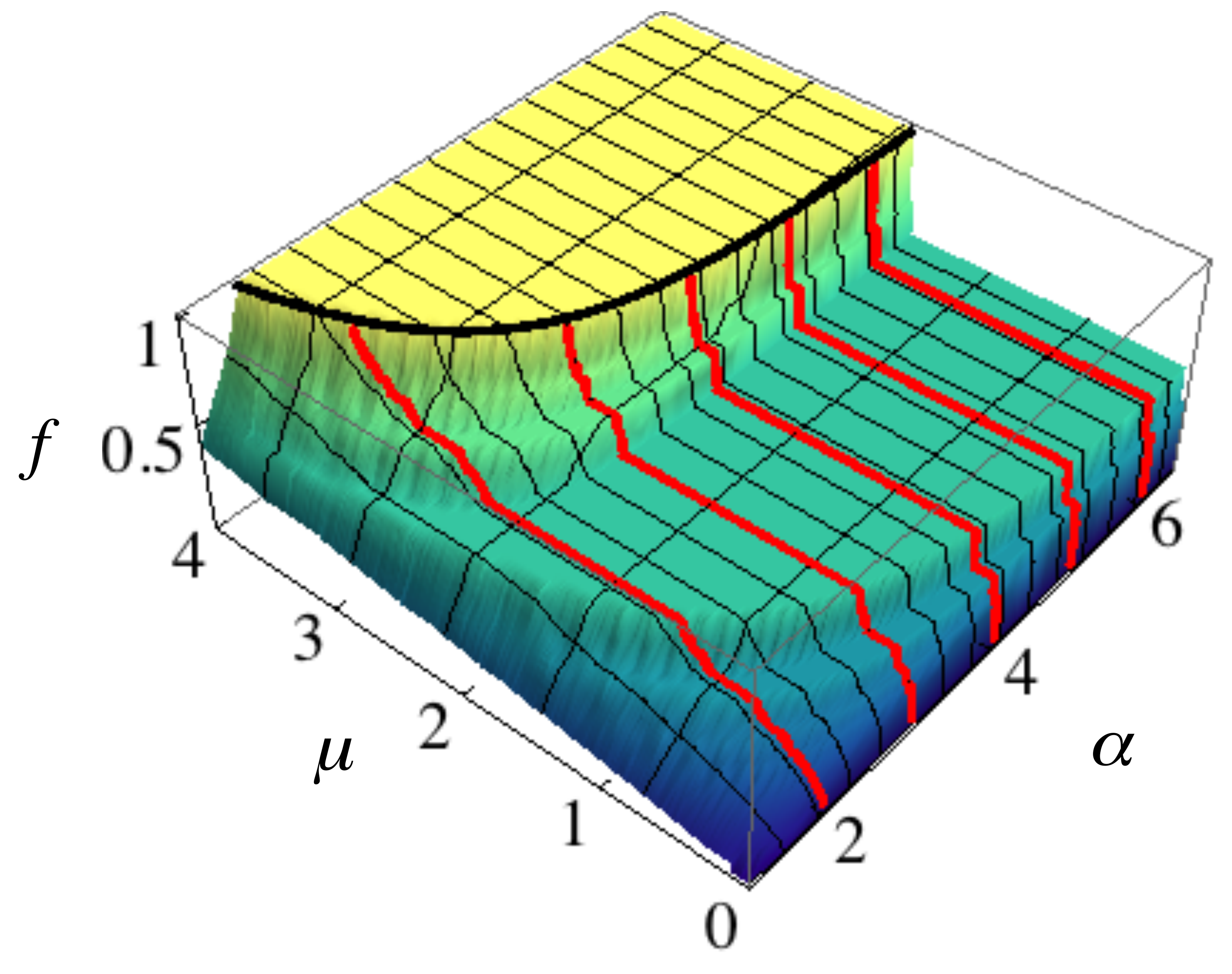}
	\caption{Ground state filling fraction $f$ of Hamiltonian (\ref{1bHp}) as functions of the chemical potential $\mu$ and the power $\alpha$ of the power-law interaction potential. The large plateau at half-filling, $f=1/2$, corresponds to the configuration of  $101010\cdots$. The red solid lines are analytical results obtained from Eqs. (\ref{mu1}) and (\ref{mu2}) at  $\alpha=2,3,4,5$, and $6$. The black line is the critical chemical potential $\mu_c(\alpha)=2\zeta(\alpha)$ at which the ground states of Hamiltonian (\ref{1bHp}) turn into the fully-filled particle states with $f=1$. }
	\label{fig1b}
\end{figure}

We numerically obtain the staircase structure by varying both the power $\alpha$ of the repulsive power-law interaction and the chemical potential. The result is shown in Fig.~\ref{fig1b}, which has a ``devil's terrace" structure. The big plateau at filling fraction $f=1/2$ corresponds to a configuration of $101010\cdots$. Its width increases as $\alpha$ increases since the large commensurate phases (with large $p$) occupy negligible parameter space of $\mu$ due to the fast decaying property of $1/r^{\alpha}$ at large $\alpha$. At small $\alpha$, the large commensurate phases play important roles and occupy a large portion of the parameter space of $\mu$.

When $f\leq1/2$, particles in the lattice are all separated from each other by empty sites and there is no cluster behaviour of the particles. However, when $f>1/2$, particles will cluster together to form different kinds of n-mers ($n\geq2$). In this regime, it becomes convenient to instead use the hole Hamiltonian (\ref{1bHh}). The hole sector at $f\geq 1/2$ is trivially related to the particle sector at $f\leq 1/2$ as the staircases (e.g., the red lines of Fig.~\ref{fig1b}) is symmetric around $f=1/2$ along the $\mu$ direction\cite{staircase82}. When the chemical potential of the particles is zero, the ground state configuration would have no particles in it and similarly, if the chemical potential of the holes is zero, one would have no holes in the lattice, i.e., the transition point to a fully-filled particle state with $f=1$ can be obtained by setting the hole chemical potential to zero. This allows us to derive the critical chemical potential analytically, $\mu_c(\alpha)=2\zeta(\alpha)$. The analytical result (marked by the black curve in Fig.~\ref{fig1b}) agrees with the numerical calculation. In the same figure, we also present the analytical result from Eqs. (\ref{mu1}) and (\ref{mu2}) at $\alpha=2,3,4,5$, and $6$ on top of the numerical data in red lines, which agree with each other very well.

\section{Two-body interaction driven staircases without particle-hole symmetry}
\label{sec5}

In this section we turn to the discussion of devil's staircases corresponding to the ground state of Hamiltonian (\ref{2bHp}). We will mainly focus on two aspects of the problem. First, we would like to understand how the range of the attraction $R$ changes the structure of the devil's staircases. Second, we investigate the effect of the power $\alpha$ of the interaction potential $V(r)=(R+1)^{\alpha}/r^{\alpha}$. For simplicity, we will consider the case where the short-range attraction $W(i)$ ($i=1,\cdots, R$) are the same and equal to $W$. 

In particular, we find that the feature of the staircase nontrivially depends on $\alpha$. For certain $\alpha$ the staircase can be described by a union of two pure sub-staircases, i.e., a pure $(R+1)$-mer particle sub-staircase and a pure $(R+1)$-mer or $(R+2)$-mer hole sub-staircase. For other values of $\alpha$, the staircases consist of more than two kinds of basic building blocks.  

We use an ``complexity parameter" $\mathcal{P}$ to describe this effect. When the value of $\alpha$ is such that the whole staircase can be described by two pure sub-staircases (a pure $n$-mer particle sub-staircase and a pure $m$-mer hole sub-staircase), i.e., a single pair of integers $(n,m)$ is sufficient to describe the emergent staircase, then $\mathcal{P}(\alpha)=1$, otherwise $\mathcal{P}(\alpha)=0$ which indicates the emergence of more complicated structures. In the following, we will study the staircases of Hamiltonian (\ref{2bHp}) by considering $R=1$ and $R=2$. A general description of $R\geq 3$ based on the data at $R=3,4,5$, and $6$ will also be presented.
\begin{figure}[htbp]
\centering
\includegraphics[width=0.85\columnwidth]{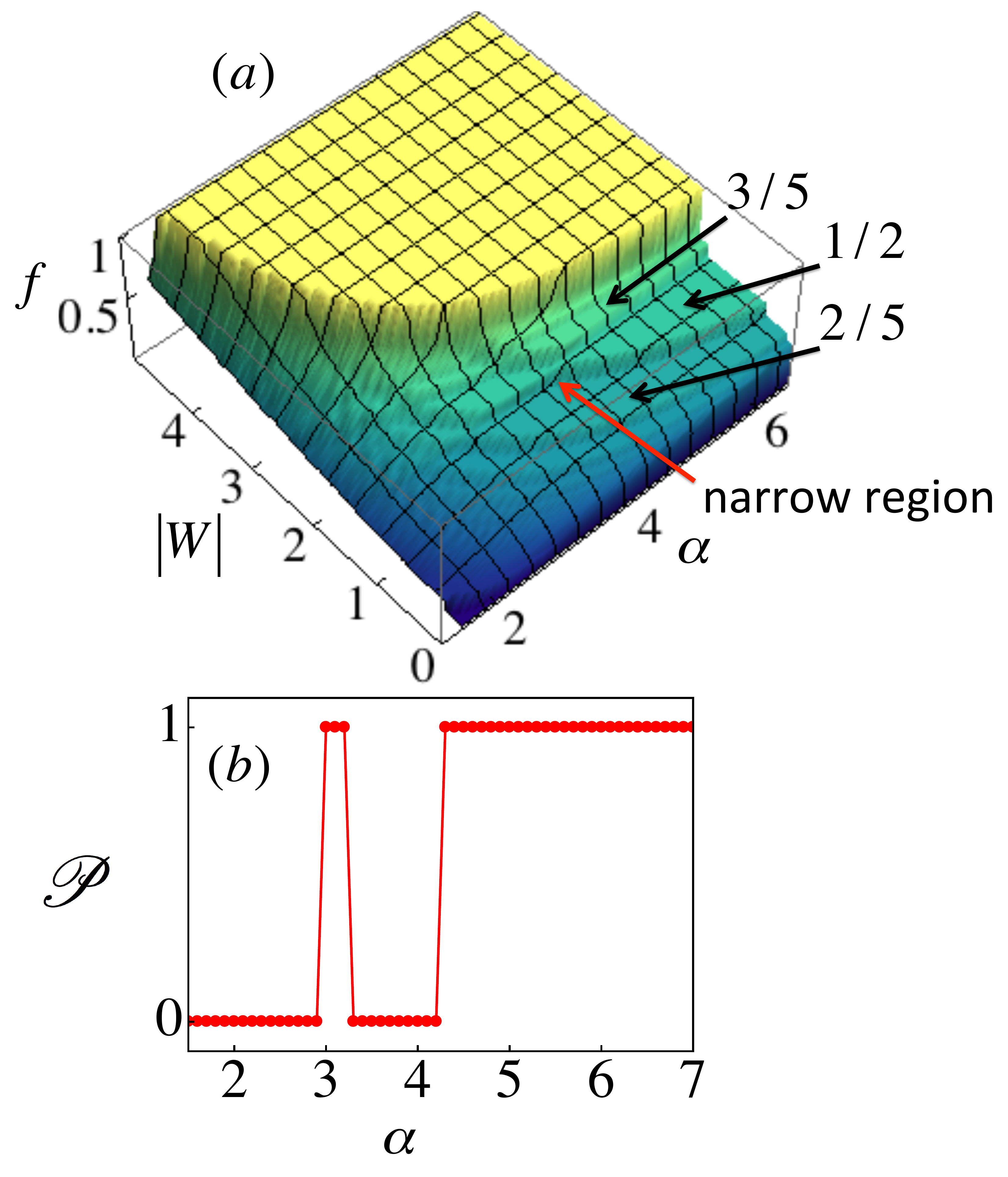}
\caption{(a) Ground state filling fraction $f$ of Hamiltonian (\ref{2bHp}) at $R=1$ as functions of the nearest-neighbour attraction $W$ and the power $\alpha$ of the long-range tail of the potential. (b) The complexity parameter $\mathcal{P}(\alpha)$. Only when $\mathcal{P}(\alpha)=1$, the staircase is described by a union of two pure sub-staircases. Dots are numerical data and the line is used to guide the eye.}
	\label{fig2bR1}
\end{figure}

\subsection{$R=1$}
When the range of the attraction is $R=1$, only the nearest-neighbour interaction is attractive and the interactions from next nearest neighbour onwards ($r=2,\cdots$) are repulsive and follow the form $V(r)=2^{\alpha}/r^{\alpha}$. The case of $\alpha=6$, corresponding to the van der Waals interaction, has been studied in detail by us in a recent work~\cite{staircase15PRL} based on a concrete system of Rydberg atoms. We found that the staircase structure has a dimer particle sub-staircase with $f\leq 1/2$ and a dimer hole sub-staircase with $f\geq 1/2$, as shown in Fig. \ref{fig2bR1}(a). The broken particle-hole symmetry in this case does not manifest in the different sizes of the clusters in the two sectors, but rather in the asymmetric shape of the staircase along the $W$-direction in the vicinity of $f=1/2$. There is no symmetry around $f=1/2$ and the relative width of certain plateaus can change significantly as we increase $\alpha$. For example, the plateau corresponding to $f=1/2$ ($f=2/5$) becomes very narrow (wide) around $\alpha=4$, as can be seen in Fig.~\ref{fig2bR1}. Such feature is not found in the Ising model studied by Bak and Bruinsma.
\begin{figure}[htbp]
\centering
\includegraphics[width=1\columnwidth]{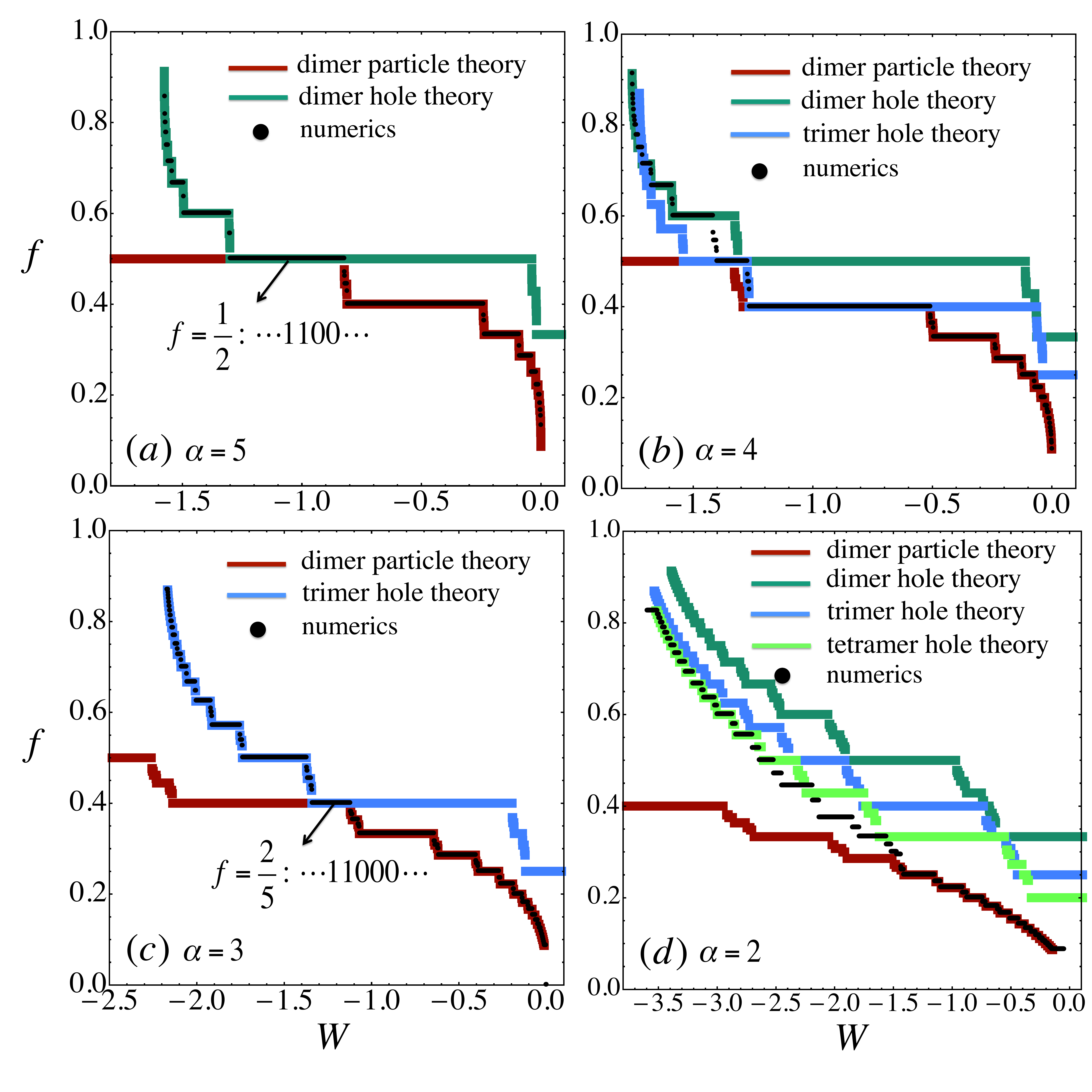}
\caption{Comparison between analytically and numerically calculated staircases at $\alpha=2,3,4,5$ as shown in Fig. \ref{fig2bR1}. At $\alpha=3,5$ the staircase can be described by a union of two pure sub-staircases in both the particle and the hole sectors with $\mathcal{P}=1$. The case of $\alpha=5$ is similar to that of our previous work \cite{staircase15PRL} where we considered $\alpha=6$. At  $\alpha=3$, we find a dimer particle sub-staircase and a trimer hole sub-staircase which meet at $f=2/5$ with the ground state configuration of $1100011000\cdots$. At $\alpha=2,4$, the staircases at the hole sector contain n-mer holes of different kinds.}
\label{fig2bR1d}
\end{figure}

Our next goal is to understand the dependence of the staircases on the interaction exponent $\alpha$. The complexity parameter $\mathcal{P}(\alpha)$ shows two regions where the staircases can be described by two pure sub-staircases. One region is $\alpha >4.2$ and another is around $\alpha=3$ as shown in Fig. \ref{fig2bR1}(b). When $\alpha >4.2$, the staircase can be described by two pure sub-staircases, i.e., a dimer particle sub-staircase at $f\leq 1/2$ and a dimer hole sub-staircase at $f\geq 1/2$ (a representative case of $\alpha=5$ is shown in Fig.~\ref{fig2bR1d}(a)). The two sub-staircases meet at $f=1/2$ with a configuration of $11001100\cdots$. Decreasing $\alpha$ in this regime narrows this central plateau (at $f=1/2$) up to $\alpha \approx 4.2$ where it disappears.

Around $\alpha=3$, we find the staircase consists of a dimer particle sub-staircase in the sector of $f\leq 2/5$ and a trimer hole sub-staircase in the sector of $f\geq 2/5$ (see Fig.~\ref{fig2bR1d}(c) for an example with $\alpha=3$). This is intriguing, and is a new manifestation of the particle-hole symmetry breaking. Decreasing $\alpha$ in this regime the central plateau at $f=2/5$ become narrow up to the point where it completely disappears. As shown in the figure, we do not find other regimes where the staircase can be described by two pure sub-staircases. 

When the hole sector can be described by a single kind of cluster hole a simple analytic calculation of the phase boundary is possible. For example, in the regime of $\alpha>4.2$, the hole sector can be solely described by dimer holes and one can find the exact transition point to the fully filled $f=1$ particle state by setting the dimer hole chemical potential  [Eq.(\ref{EqR12})] to zero, which leads to,
 \begin{gather}
 W_c=-\frac{4}{3}\sum_{r=2}^{\infty}V(r)=-\frac{4[\zeta (\alpha)-1]2^{\alpha}}{3}.
 \end{gather}
Similarly, for the regime around $\alpha =3$, the hole sector can be described by trimer holes. From Eq.~(\ref{EqR13}) we obtain then the transition point,
 \begin{gather}
 W_c=-\frac{6\sum_{r=2}^{\infty}V(r)-V(2)}{4}=-\frac{6[\zeta (\alpha)-1]2^{\alpha}-1}{4}.
 \end{gather}

For other values of $\alpha$, the staircase has more complicated structures. Examples with $\alpha=2$ and  $4$ are shown in Figs.~\ref{fig2bR1d}(b) and \ref{fig2bR1d}(d) where one cannot describe the emergent staircase with a union of two pure sub-staircases. For example, we list the ground state configurations at $\alpha=4$ for two different $W$,
\begin{gather}
W=-1.27 \rightarrow 11100011000 \cdots f=\frac{5}{11},\\
W=-1.41 \rightarrow 11100111000  \cdots  f=\frac{6}{11},
\end{gather}
which clearly shows that the staircase consists of dimer particles, trimer particles and dimer holes, and trimer holes.

%+++++++++++++++++++++++++++++++

\subsection{$R=2$}

\begin{figure}[htbp]
\centering
\includegraphics[width=0.75\columnwidth]{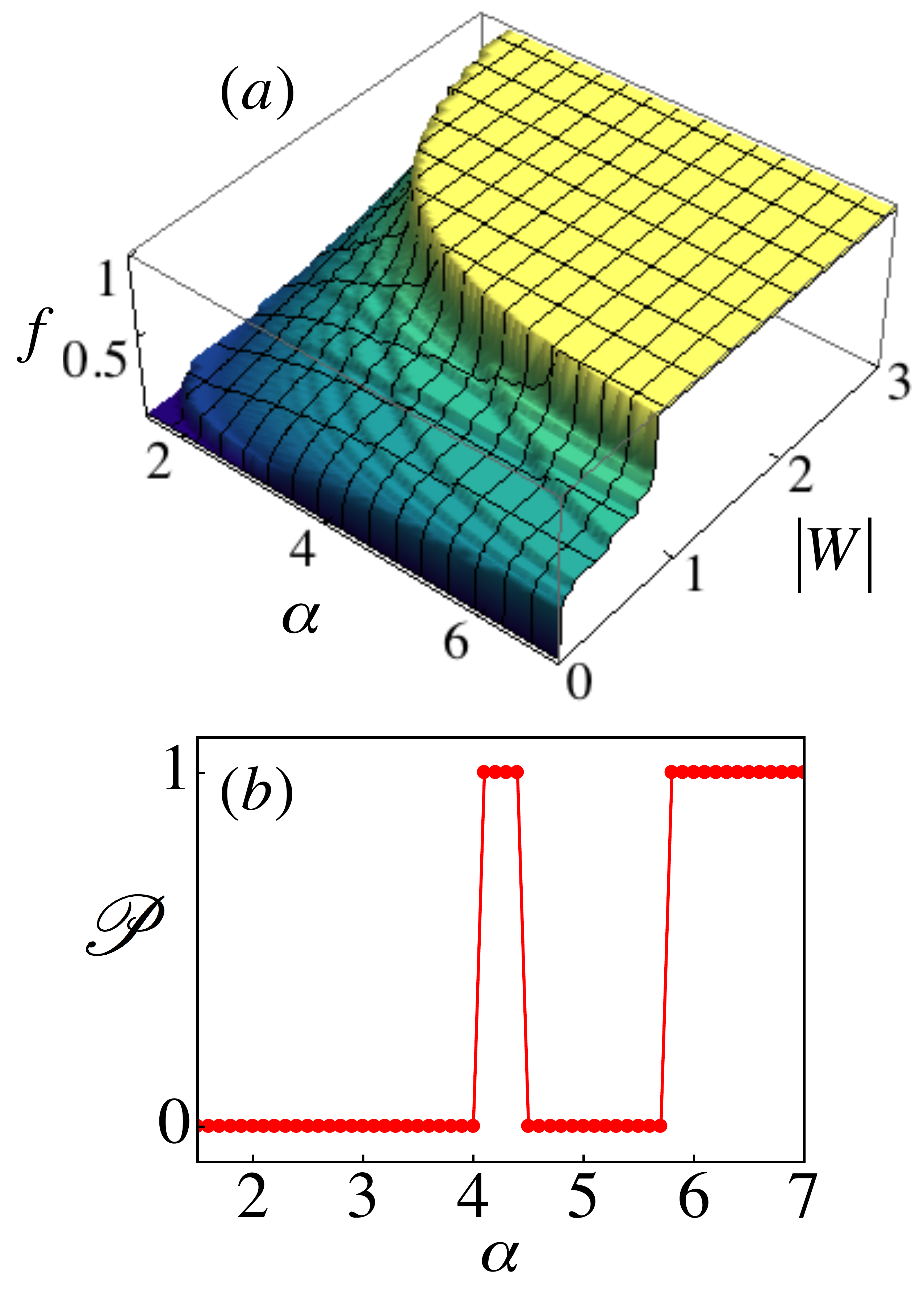}
\caption{(a) Ground state filling fraction $f$ for $R=2$ as functions of the attractive interaction strength $W$ and the power $\alpha$. (b) The complexity parameter  $\mathcal{P}(\alpha)$. See also Fig.~\ref{fig2bR1}.}
\label{fig2bR2}
\end{figure}

We will now consider the case $R=2$, i.e. where the attractive range of the interaction potential spans two sites. For simplicity, we will focus on the case where $W(1)=W(2)=W<0$. The long-range repulsive interaction tail now becomes $V(r)=3^{\alpha}/r^{\alpha}$ ($r=3,\cdots$). For such interactions, three particles tend to cluster together on neighbouring lattice sites, to form a trimer particle serving as the basic building block of the staircase at the particle sector. The ground state filling fraction $f$ of Hamiltonian (\ref{2bHp}) at $R=2$ is shown in Fig. \ref{fig2bR2}(a).

\begin{figure}[htbp]
\centering
\includegraphics[width=1\columnwidth]{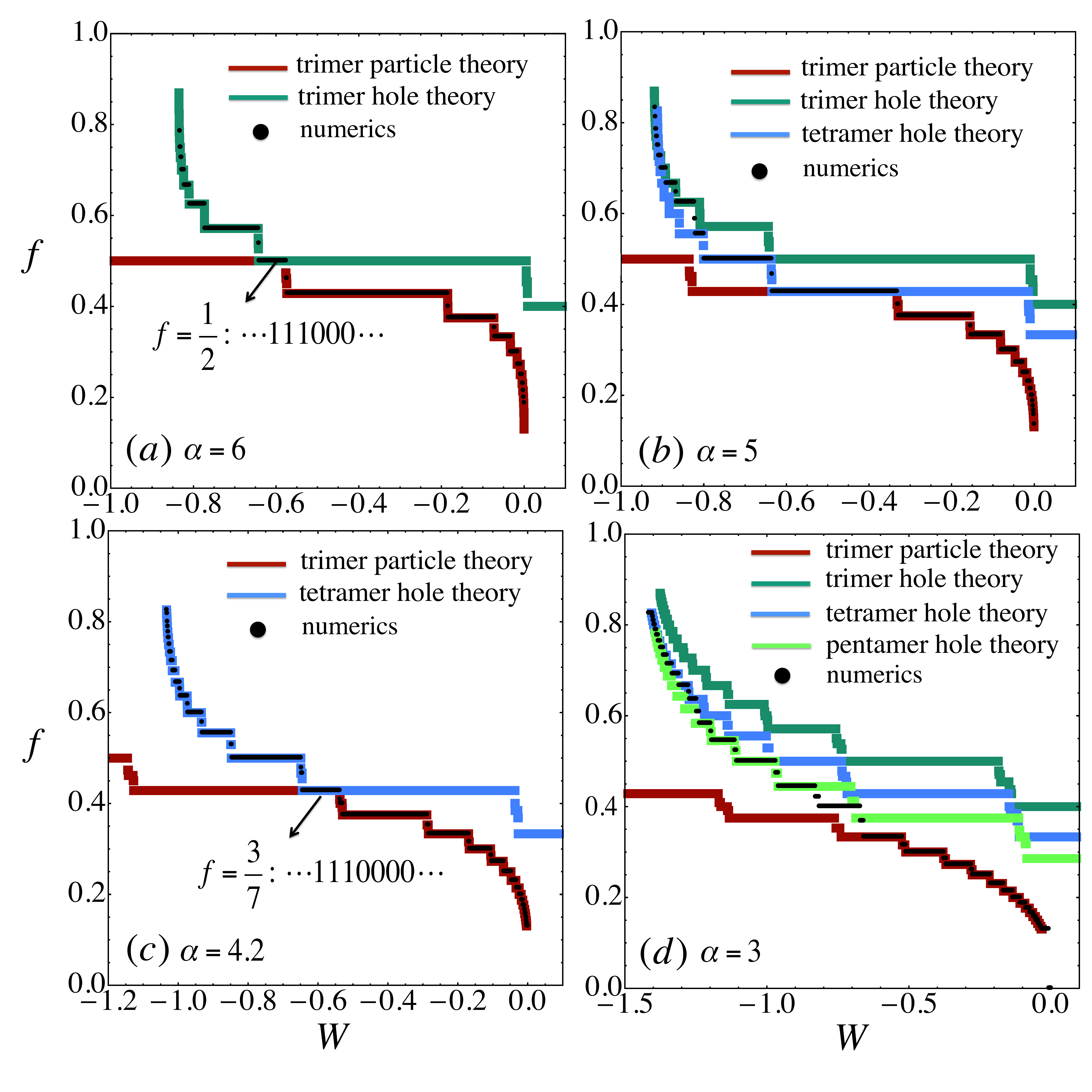}
\caption{Analytically calculated staircases compared to the numerically obtained staircases at $\alpha= 6,5,4.2$, and $3$ in panels (a-d) respectively. The other parameters are $R=2$ and $W(1)=W(2)=W$. See also Fig.~\ref{fig2bR1d}. }
\label{fig2bR2d}
\end{figure}

There are again two regimes where the staircase can be described by a union of two pure sub-staircases both in the particle and hole sectors, as shown by the complexity parameter $\mathcal{P}(\alpha)$ in Fig.~\ref{fig2bR2}(b). When $\alpha>5.8$, the staircase can be described by a union of trimer particle sub-staircase in the particle sector and a trimer hole sub-staircase in the hole sector, where the two sub-staircases meet at $f=1/2$ with a configuration $111000111000\cdots$. Around $\alpha=4.2$, we obtain a trimer particle sub-staircase in the particle sector and a tetramer hole sub-staircase in the hole sector, where the two sub-staircases meet at $f=3/7$ with a configuration $11100001110000\cdots$. Apart from these two regimes, the staircases cannot be described as a union of two pure sub-staircases. The detailed results of the staircases at $\alpha=3,4.2,5$, and $6$ are presented in Fig. \ref{fig2bR2d}.

We can also find the exact transition points to the unit filling $f=1$ particle states in the two regimes where the staircases can be described by a single kind of cluster hole in the hole sector. This is done by setting the trimer hole and tetramer hole chemical potentials of Eqs.~(\ref{EqR22}) and (\ref{EqR23}) to zero, which yields
\begin{gather}
W_c=-\frac{2}{3}\sum_{r=3}^{\infty}V(r)=-\frac{2[\zeta (\alpha)-1-\frac{1}{2^{\alpha}}]3^{\alpha}}{3},
\end{gather}
and
\begin{gather}
W_c=-\frac{8\sum_{r=3}^{\infty}V(r)-V(3)}{11}=-\frac{8[\zeta (\alpha)-1-\frac{1}{2^{\alpha}}]3^{\alpha}}{11}.
\end{gather}

%+++++++++++++++++++++++++++++++

\subsection{$R>2$}
\begin{figure*}[htbp]
\centering
\includegraphics[width=1.8\columnwidth]{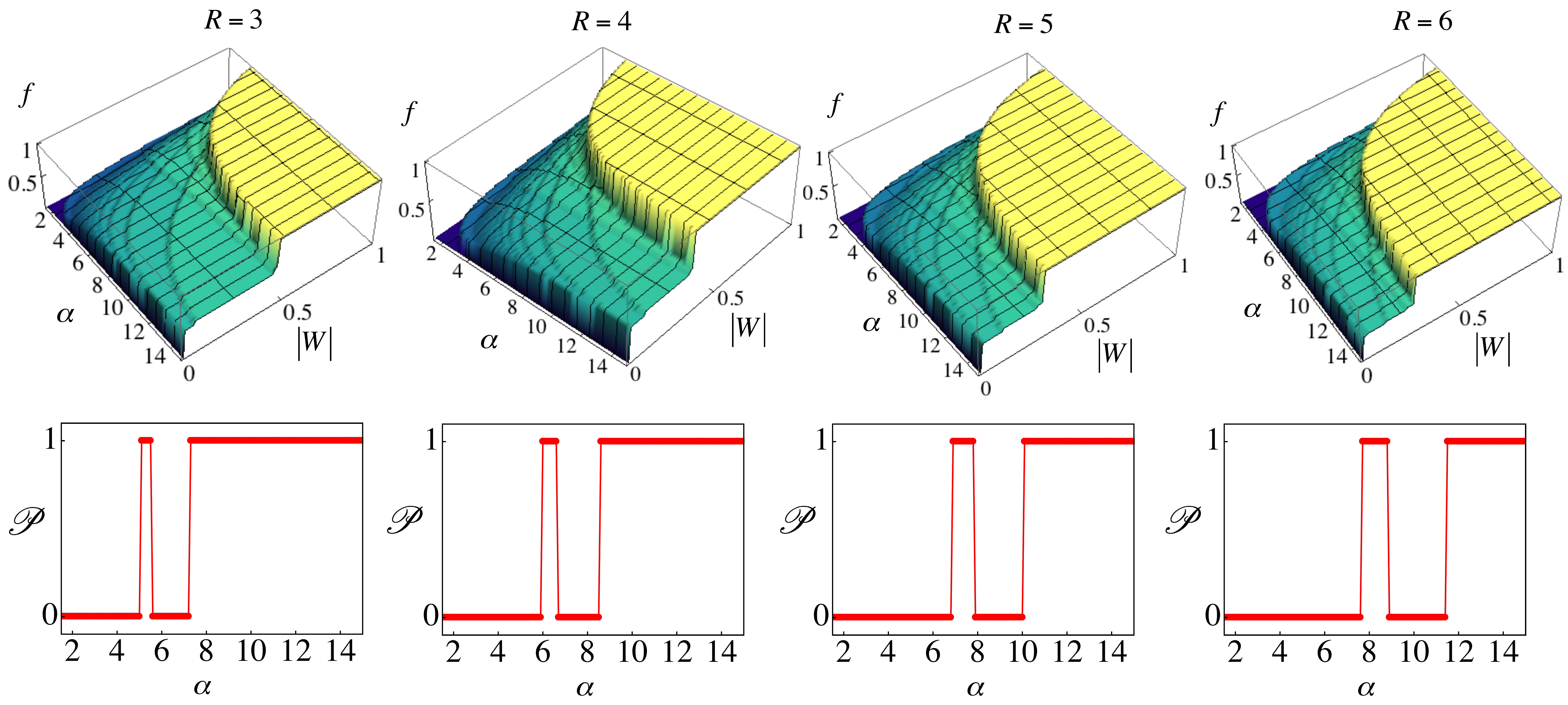}
\caption{(a) Ground state filling fraction $f$ of Hamiltonian (\ref{2bHp}) for $R=3,4,5,6$ shown as a function of the attractive interaction strength $W(i)=W$ ($i=1,2,\cdots, R$) and the power $\alpha$ of the long-range power law repulsion $V(r)=(R+1)^{\alpha}/r^{\alpha}$ with ($r=R+1, \cdots$). Shown in (b) is $\mathcal{P}(\alpha)$ similar to Figs. \ref{fig2bR1}(b), \ref{fig2bR2}(b) at $R=1$ and $R=2$. }
\label{fig2bR3456}
\end{figure*}

When $R>2$, the qualitative feature of the physics is largely similar to the case $R=1$ and $R=2$. The sub-staircase in the particle sector is built up from $(R+1)$-mer particles. In the hole sector, there are two regimes where the staircase can be described by a union of two pure sub-staircases in both the particle and the hole sectors. In one region, we have a  $(R+1)$-mer particle sub-staircase and a $(R+1)$-mer hole sub-staircase, where the two sub-staircases meet at $f=1/2$ with a configuration of $1\cdots10\cdots0\cdots$ (with both $R+1$ ``1"s and $R+1$ ``0"s). There is also a narrow region of $\alpha$, where the staircase is made of a $(R+1)$-mer particle sub-staircase and a $(R+2)$-mer hole sub-staircase, where the two sub-staircases meet at $f=(R+1)/(2R+3)$ with a configuration of $1\cdots10\cdots0\cdots$ (i.e., $R+1$ particles and $R+2$ holes). Apart from these two regimes, the staircases cannot be described by a union of two pure sub-staircases. These features can be seen from the ground state filling fraction $f$ and order parameter $\mathcal{P}(\alpha)$ of Hamiltonian (\ref{2bHp}) with $R=3,4,5$, and $6$ as shown in Fig. \ref{fig2bR3456}.  Moreover, the critical transition points of the $(R+1)$-mer hole sub-staircase and the $(R+2)$-mer hole sub-staircase to the full-filled $f=1$ particle states can also be found from Eq. (\ref{CPnHole}).

 %%%%%%%%%%%%%%%%%%%%%%%%%%%%%%%%%%%%%%%%

\section {$N (\geq3)$-body interaction driven staircase}
\label{sec6}
The above results indicate that there should be $R$-mer staircase when the range of the attractive interactions is $R-1$. A natural question is whether such $R$-mer staircase can be induced by $R$-body interactions directly? To answer this question, we investigate the following model Hamiltonian, which contains a two-body long-range repulsive interaction described by $V(r)$, and a $N\geq3$-body attractive interaction by $U_N$,
\begin{gather}
H=\sum_{i=-\infty}^{\infty}  \sum_{r=1}^{\infty}V(r) n_i n_{i+r}-U_N \sum_{i=-\infty}^{\infty} n_i n_{i+1} \cdots n_{i+N-1}.
\label{nbH}
\end{gather}

Numerical calculations of the above Hamiltonian show that in the ground state, there is always a direct transition from the empty state of $\cdots 000 \cdots $ to the fully filled state of $\cdots 111 \cdots $. The energies of the two states are 0 and $\sum_{r\geq 1}V(r)-U_N$. Hence the transition happens when $(\sum_{r\geq 1}V(r)-U_N )<0$, i.e., $U_N>\sum_{r\geq 1}V(r)=\zeta(\alpha)$.

In the following, we provide a simple explanation of this result based on energy arguments. The energy of a $N$-mer is
\begin{gather}
E_N=(N-1)V(1)+(N-2)V(2)+\cdots V(N-1)-U_N. \nonumber
\end{gather}
One can readily show that the energy of two separate $N$-mers are $2E_N+E_{\text{int}}$ with $E_{\text{int}}$ the interaction energy of the two $N$-mers, which is larger than the energy of a $(N+1)$-mer,
\begin{eqnarray}
2E_N&&+E_{\text{int}}-E_{N+1}=\nonumber \\&&(N-2)V(1)+(N-3)V(2)+\cdots +(-1)V(N)+E_{\text{int}}>0,\nonumber
\end{eqnarray}
when $N\geq 3$. This result excludes the possibility of having exotic staircases driven solely by N-body attraction when $N\geq3$. 

 %%%%%%%%%%%%%%%%%%%%%%%%%%%%%%%%%%%%%%%%

\section{Discussion and outlook}
\label{sec7}

We have explored a new class of devil's staircases that exhibit a broken particle-hole symmetry. The symmetry breaking is purely induced by the interplay between short-range attraction and long-range repulsion. When the staircase can be described by a union of two pure sub-staircases in both the particle and hole sectors, the value of the critical attractive strength $W_c$ and the ``width" of the stability region can be found analytically. These confirm that the resulting staircase is complete \cite{staircase82, staircase15PRL}. However, when the staircase contains mixtures of n-mers of different kinds, it is an open question whether an analytic understanding of the staircase structure can be obtained. Another possible way to understand the problem-which may lead to an answer-is to consider periodic configurations as consisting of segments of different phases separated by interfaces \cite{interface}, where the nature of interface interactions determine the detailed structure of the phase diagram. One interesting question is why for attractive interactions with range $R$, there can only be $(R+1)$-mer hole and $(R+2)$-mer hole sub-staircases but not $(R+3)$ hole sub-staircase in the hole sector for power law repulsion. The answer might be that without particle-hole symmetry, the staircase structure will depend on the specific form of the repulsive tail itself. This also suggests an interesting way to manipulate the hole part of the staircase by controlling the form of the repulsive tail. We expect that for example with exponential interactions \cite{staircase_exponential} for the repulsive part, the hole part may indeed display a different structure. Furthermore, it is known that some 2D lattice gas \cite{2D_staircase_lattice_gas} and adsorption \cite{2D_staircase_dimer_adsorption} models can have devil's staircase of phase transitions in the ground state. So it would be interesting to extend the current work to 2D by coupling 1D chains transversely. 

A further interesting problem for future studies is the exploration of the role of thermal and quantum fluctuations. For example, for the ANNNI model, it is the thermal fluctuations that stabilize the staircase. Quantum fluctuations, however, can destroy the staircase at zero temperature \cite{quantum_fluctuation_staircase, staircase15PRL, two_component_staircase}, i.e., the stability regions shrink and at most a finite number of commensurate phases survives. So it would be interesting to understand how quantum fluctuations will melt the emerging hybrid staircases, such as the dimer-particle and trimer-hole staircase [see Fig.~\ref{fig2bR1d}(c)], studied in this paper. One might be able to address these questions experimentally for example with a recently established quantum simulator platform based on Rydberg atoms~\cite{RydCry_science, Browaeys_nature, Lukin_2017,review_sweep}. The preparation of the ground state of our model on a Rydberg atom quantum simulator requires an adiabatic sweep protocol. A detailed discussion of this procedure can be found in the recent review \cite{review_sweep}.

\section{ Acknowledgments}
We thank Emanuele Levi and  Ji\v{r}\'{i} Min\'{a}\v{r} for their contribution at the early stage of this work.  
The research leading to these results has received funding from the
European Research Council under the European Union's Seventh Framework
Programme (FP/2007-2013) / ERC Grant Agreement No. 335266 (ESCQUMA), the
EU-FET Grant No. 512862 (HAIRS), the H2020-FETPROACT-2014 Grant No.
640378 (RYSQ), EPSRC Grant No. EP/M014266/1, and the UKIERI-UGC Thematic Partnership No. IND/CONT/G/16-17/73. I.L. gratefully  acknowledges  funding  through  the  Royal Society
Wolfson Research Merit Award.

%%%%%%%%%%%%%%%%%%%%%%%%%%%%%%%%%%%%%

\appendix
\begin{figure*}[htbp]
	\centering
	\includegraphics[width=1.9\columnwidth]{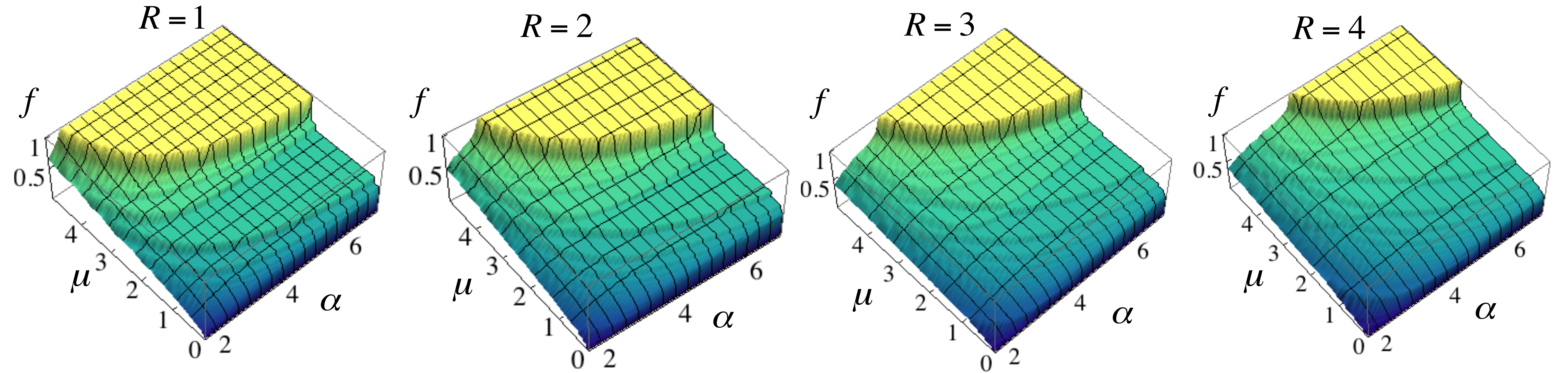}
	\caption{Ground state filling fraction $f$ of Hamiltonian (\ref{2bHh}) for $R=1,2,3,4$ as functions of the chemical potential $\mu$ and the power $\alpha$ of the long-range power-law repulsion $V(r)=R^{\alpha}/r^{\alpha}$ ($r=R+1, \cdots$), where the short-range attractions $W(i)$ ($i=1,2,\cdots, R$) have been set to zero. This chemical potential $\mu$ driven staircase has the particle-hole symmetry, so the hole sector is trivially related to the particle sector, which is very different from our two-body attraction $W(i)$ driven staircase, where the hole sector contains very rich physics as studied in the main text.}
	\label{polymer}
\end{figure*}

\section{Stability regions}
\label{secA}

In this part of the appendix, we give a brief derivation of the stability regions of Eqs. (\ref{mu1}) and (\ref{mu2}) used in the main text (see Refs. \cite{Hubbard, Uimin} for the original literature). The energy of the ground state configuration can be written as
\begin{eqnarray*}
E_0=E_1+E_2+\cdots+E_n+\cdots+E_{\infty}+E_{\mu},
\end{eqnarray*}
where $E_{1,2,\cdots,\infty}$ is the interaction energy with nearest-neighbour and next-neast-neighbour and so on and $E_{\mu}$ is the energy with the chemical potential term.
For any filling fraction of $f=\frac{q}{p}$, the $n$-nearest-neighbour interaction energy of the most homogeneous configuration requires $np=r_n q+\alpha_n$, where $0\leq\alpha_n<q$. To make this relation clear, we rewrite it as $np=r_n x+(r_n+1)(q-x)$ by introducing a new integer $x=(r_n+1)q-np$. It means that there are $x$ particles separated from each other by $r_n$ lattice sites while $q-x$ particles separated from each other by $r_n+1$ lattice sites. The energy of $E_n$ with $L$ periods (a very large number) is then
\begin{eqnarray*}
E_n=[xV(r_n)+(q-x)V(r_n+1)]L.
\end{eqnarray*}

Now if we have one more particle in the above configuration, the interaction will reorganise the particle distribution such that $npL=r_n y+(qL+1-y)(r_n+1)$, i.e., compared with the above case, we have $qL+1$ particles, from which we can get $y=(qL+1)(r_n+1)-npL$, so the energy of $E_n^{+}$ with one more particle is
\begin{eqnarray*}
E_n^{+}=yV(r_n)+(qL+1-y)V(r_n+1).
\end{eqnarray*}

In the same way for one less particle in the configuration, $npL=r_n z+(qL-1-z)(r_n+1)$ and we get $z=(qL-1)(r_n+1)-npL$ and
\begin{eqnarray*}
E_n^{-}=zV(r_n)+(qL+1-z)V(r_n+1).
\end{eqnarray*}

However, if $\alpha_n=0$, i.e, $x=q$, we have slightly different situations,
\begin{eqnarray*}
&&E_n=[qV(r_n)]L\nonumber\\
&&E_n^{+}=yV(r_n)+(qL+1-y)V(r_n-1) \nonumber\\
&&E_n^{-}=zV(r_n)+(qL-1-z)V(r_n+1) \nonumber\\
&&r_ny+(qL+1-y)(r_n-1)=npL\Rightarrow y=npL-(qL+1)(r_n-1)\nonumber\\
&&r_nz+(qL-1-z)(r_n+1)=npL\Rightarrow z=(qL-1)(r_n+1)-npL\nonumber
\end{eqnarray*}

In summary, we get
\begin{eqnarray}
\mu^{+}&&=\sum_{n=1, \alpha_n\neq 0}^{\infty} (E_n^{+}-E_n)=\sum_{n=1, \alpha_n\neq 0}^{\infty} [(r_n+1)V(r_n)-r_nV(r_n+1)] \nonumber \\
&&+\sum_{n, \alpha_n=0}^{\infty}[(-r_n+1)V(r_n)+r_nV(r_n-1)]\nonumber
\end{eqnarray}
and
\begin{eqnarray*}
\mu^{-}&&=\sum_{n=1, \alpha_n\neq 0}^{\infty} (E_n-E_n^{-})=\sum_{n=1, \alpha_n\neq 0}^{\infty} [(r_n+1)V(r_n)-r_nV(r_n+1)] \nonumber \\
&&+\sum_{n, \alpha_n=0}^{\infty}[(r_n+1)V(r_n)-r_n V(r_n+1)]
\end{eqnarray*}
which are used in the main text.

%%%%%%%%%%%%%%%%%%%%%%%%%%%

\section{Polymer staircases with particle-hole symmetry}
\label{secB}

In the main text, we study the devil's staircase physics described by Hamiltonian (\ref{2bHp}) where the short-range two-body attraction is the main driving force for the emergence of devil's staircase behaviour. The emergent staircases could be termed as polymer staircases, which consist of some basic clusters with different sizes. Our main motivation of the main text is to study the effect of broken particle-hole symmetry on the staircase structure where the polymer behaviour of the staircases is a byproduct. However, we note that in the literature there have been studies where the motivation was to look for mechanism to form a polymer staircase, whereas the particle-hole symmetry is not the focus. Actually, one can still preserve the particle-hole symmetry of the polymer staircases by using the single-body chemical potential as the driven mechanism, e.g., the papers by Jedrzejewski and Miekisz \cite{nonconvex_EPL, nonconvex_long} fit to this category.

Here we would like to briefly discuss the connection of our work with those of Jedrzejewski and Miekisz~\cite{nonconvex_EPL, nonconvex_long}. These authors have proven rigorously the existence of the dimer staircases in 1D lattice gas models with certain nonconvex long-range interactions, where the particle density versus the chemical potential, $\rho(\mu)$, exhibits the complete devil's staircase structure. The authors also speculated that for interactions with values near zero for distances up to $R$ and strictly convex from distance $R+1$ onwards, the ground state forms a $R+1$-mer staircase. We present in Fig. \ref{polymer} the staircase structure according to the speculation of the papers by Jedrzejewski and Miekisz \cite{nonconvex_EPL, nonconvex_long} using the numerical tool described in our main text, where we set $W(1)=\cdots W(R)=0$, and from distance $R+1$ onwards, the repulsive interaction has the form of $(R+1)^{\alpha}/r^{\alpha}$. From the numerical results, we verify that given $R$, the staircase is a $(R+1)$-mer staircase with exact particle-hole symmetry (see Fig. \ref{polymer}, where the mesa is symmetric with respect to the $f=1/2$ plateaus possessing a configuration of $(1\cdots1 0\cdots 0) \cdots$  with both $R+1$ ``1"s and $R+1$ ``0"s in one period). Due to the particle-hole symmetry, the hole sectors of these staircases are trivially related to the particle sectors of them. This kind of staircase bears some similarities with both the traditional staircases presented in section \ref{sec4} and our two-body attraction driven staircases studied in section \ref{sec5} in the sense that they have particle-hole symmetry as the traditional staircases but show cluster behaviour as our two-body attraction driven staircases.

%%%%%%%%%%%%%%%%%%%%%%%%%%%

\end{document}